*The Massalia asteroid family as the origin of ordinary L chondrites*

M. Marsset[1,2], P. Vernazza[3], M. Brož[4], C. A. Thomas[5], F. E. DeMeo[2], B. Burt[6], R. P. Binzel[2], V. Reddy[7], A. McGraw[7], C. Avdellidou[8], B. Carry[8], S. M. Slivan[2], D. Polishook[9]

[1] European Southern Observatory (ESO), Alonso de Cordova 3107, 1900 Casilla Vitacura, Santiago, Chile
[2] Department of Earth, Atmospheric and Planetary Sciences, Massachusetts Institute of Technology, 77 Massachusetts Avenue, Cambridge, MA 02139, USA
[3] Aix Marseille Univ, CNRS, CNES LAM, Institut Origines, Marseille, France
[4] Charles University, Faculty of Mathematics and Physics, V Holešovičkách 2, 18000 Prague, Czech Republic
[5] Department of Astronomy and Planetary Science, Northern Arizona University, PO Box 6010, Flagstaff, AZ 86005, USA
[6] Lowell Observatory, 1400 W. Mars Hill Road, Flagstaff, AZ, 86001, USA
[7] Lunar and Planetary Laboratory, University of Arizona, Tucson, AZ 85721, USA
[8] Université Côte d'Azur, CNRS - Lagrange, Observatoire de la Côte d'Azur, CS 34229, 06304, Nice Cedex 4, France
[9] Faculty of Physics, Weizmann Institute of Science, 234 Herzl St., Rehovot 7610001 Israel.

**Studies of micrometeorites in mid-Ordovician limestones and Earth's impact craters indicate that our planet witnessed a massive infall of ordinary L chondrite material ~466 million years (My) ago (Heck et al. 2017, Schmieder & Kring 2020, Kenkmann 2021) that may have been at the origin of the first major mass extinction event (Schmitz et al. 2019). The breakup of a large asteroid in the main belt is the likely cause of this massive infall. In modern times, material originating from this breakup still dominates meteorite falls (>20% of all falls) (Swindle et al. 2014). Here, we provide spectroscopic observations and dynamical evidence that the Massalia collisional family is the only plausible source of this catastrophic event and of the most abundant class of meteorites falling on Earth today. It is suitably located in the inner belt, at low-inclination orbits, which corresponds to the observed distribution of L-chondrite-like near-Earth objects (NEOs) and of interplanetary dust concentrated at 1.4 degrees (Sykes 1990, Reach et al. 1997).**

Ordinary L chondrites are the most abundant type of meteorites recovered today. Nearly half of them are heavily shocked (e.g., Heymann 1967; Marti & Graf 1992; Rubin 1994; Bischoff et al. 2019 and references therein) and degassed, with $^{40}Ar/^{39}Ar$ radiometric ages near 470 My (e.g., Korochantseva et al. 2007; Swindle et al. 2014). This strongly suggests that an L-chondrite-like asteroid suffered a supersonic impact ~470 My ago (Haack et al. 1996), which ejected a substantial amount of material. Studies of fossil meteorites found in a 466 My old Ordovician strata in a limestone quarry in Sweden (Schmitz et al. 1997, 2001) supported these findings and further revealed that the measured abundance of fossil L chondrites implies a rate of meteoritic bombardment 1 to 3 orders of magnitude higher than at present (Heck et al. 2017; Terfelt & Schmitz 2021), with L chondrites representing ≥ 99% of all falls just after the impact. In order to explain the slow cooling rates of L chondrites, the parent body had to be larger than 100 km in diameter (Haack et al. 1996). All this implies that the impact at the origin of the L chondrite shower on Earth 466 Myrs ago must have led to the formation of a prominent asteroid family in the main belt (Greenwood et al. 2020), still visible today as a cluster in the space of proper orbital elements (semimajor axis, eccentricity and inclination; see, e.g., Nesvorný et al. 2015).

Most S-type asteroids have ordinary chondrite (OC)-like compositions (Gaffey et al. 1993, Nakamura et al. 2011). In order to identify the main source of L chondrites, past spectroscopic surveys have prioritized members in S-type families with diameters usually greater than 10 km, as they were the most accessible to observations. It follows that for families consisting of one very large remnant and the second largest remnant with $D < 10$ km (e.g., Massalia, Phocaea), spectroscopic data were only available for the largest body. These past spectroscopic surveys identified the Gefion family, located in the middle belt near the 5:2 mean-motion resonance with Jupiter, as a plausible source of L chondrites (Vernazza et al. 2014). However, subsequent simulations of the orbital evolution of the family indicated that the Gefion family is presumably older than 470 My (Aljbaae et al. 2019). Moreover, according to up-to-date catalogs, its size-frequency distribution (SFD) is shallow, which is common for relatively old families (>500 My; Brož et al. 2023).

To identify the source of the L chondrite parent body, we conducted spectroscopic observations and compiled literature data for all prominent S-type asteroid families in the main belt. Specifically, new spectroscopic observations of small family members were collected over near-infrared wavelengths (0.8–2.5 μm) using the SpeX spectrograph on NASA's InfraRed Telescope Facility (Appendix A). The telescopically measured asteroid spectra were then compared to analogous-wavelength laboratory measurements of ordinary chondrites from the RELAB database (Pieters & Hiroi 2004; Milliken et al. 2016) to identify their meteorite analog. We computed the mean spectrum of each S-type asteroid family and de-reddened them using the empirical reddening function of Brunetto et al. (2006) to account for space weathering. We then applied a radiative transfer model (Shkuratov et al. 1999) to the spectra of both asteroid and meteorite samples following the same technique as, e.g., Vernazza et al. (2014) and Binzel et al. (2019). We used the two end-member minerals olivine (ol) and a low-calcium pyroxene, namely orthopyroxene (opx), to measure their relative abundances and quantitatively evaluate mineralogy by the ratio ol/(ol+opx). The OC analog of each S-type family was then assessed by comparing the measured ol/(ol+opx) ratio of the meteorites and asteroids, and through direct comparison of their spectra (Appendix B).

Our survey mostly confirms previous findings about the mineralogy of S-type families (Vernazza et al. 2014; Table 1). There are, however, two remarkable exceptions:

Phocaea and Massalia. So far, the mineralogy of these two families was limited to that of their largest member. Here, we investigate a total number of 9 spectra of 6 family members for Phocaea, and 21 spectra of 15 family members for Massalia. The largest member of the Phocaea family, (25) Phocaea, is an LL-type; our study reveals that it is most likely an interloper in the H-dominated family. On the other hand, the largest member of the Massalia family, (20) Massalia, with ol/(ol+opx) = 0.61+/-0.02 (the error bars indicates the 1-σ scattering of values), is compatible both with H and L ordinary chondrite meteorites (Fig. 1). Our study, however, finds the other Massalia asteroid family members are L-type. Specifically, with ol/(ol+opx) = 0.67+/-0.04, the distribution of these objects is centered exactly at the location of the peak of L chondrites (Fig. 1). A genetic link between small Massalia asteroids and L chondrites is further supported by a direct comparison of their spectra (Fig. 2).

The discovery of Massalia as an L-chondrite-like family in the inner belt, in between the two most effective pathways to the near-Earth environment (the $\nu_6$ secular resonance and the 3:1 mean-motion resonance with Jupiter; Granvik et al. 2018) makes it the new prime candidate for the source of L chondrites accreted today by the Earth. The relatively young age of the family (<500 My; Nesvorný et al. 2003, Vokrouhlický et al. 2006, Spoto et al. 2015) and its center location at ~2.41 au from the Sun implies that an NEO with a size of 300 meters (roughly corresponding to the typical size of NEOs studied in near-infrared spectroscopic surveys; Marsset et al. 2022) originating from the family can only reach the Earth via the 3:1 resonance. Indeed, it would take longer than the family age, >500 My, for a 300-m asteroid to reach the $\nu_6$ through Yarkovsky drift, starting from the center of the family. On the other hand, meter-size objects can do so in only 3-10 My (Appendix C). As such, if Massalia truly is the source of shocked L chondrites, then one should expect L-chondrite-like NEOs to preferentially cluster around the 3:1 resonance with respect to the overall S-type NEO population.

In order to verify this prediction, we compiled near-infrared spectra for a total of 621 individual S-type NEOs from the MIT-Hawaii Near-Earth Object Spectroscopic (MITHNEOS) survey (Binzel et al. 2019, Marsset et al. 2022), and computed their ol/(ol+opx) ratio as performed in the case of OCs and family members. We then divided our NEO dataset into three subsets of objects with distinct ranges of mineralogy: H-chondrite-like NEOs with ol/(ol+opx) < 60%, L-chondrite-like NEOs with 62% < ol/(ol+opx) < 69% and LL-chondrite-like NEOs with ol/(ol+opx) > 72%. Bodies falling in between these ranges were not included in our analysis as being more ambiguous. Fig. 3 shows the orbital distribution of each group of objects within the space of osculating semi-major axis versus inclination and confirms a clustering of L-chondrite-like NEOs near the 3:1 resonance. We note that, on average, L-chondrite-like NEOs are found on slightly more inclined orbits (1-10 degree) with respect to Massalia (~1.4 degree), which is expected considering that NEOs are more subject to dynamical excitation through close encounters with planets, chaotic diffusion, as well as Kozai oscillations (Kozai 1962), compared to main-belt asteroids.

We further note that LL-chondrite-like NEOs cluster near the Flora family, in agreement with an origin from this family, as proposed by several authors in the past (Vernazza et al. 2008; Thomas & Binzel 2010; de León et al. 2010; Dunn et al. 2013). Additional contribution to the population of LL-chondrite-like NEOs is from the Nysa family, which we mineralogically classify for the first time in this work. We also find that the vast majority of H-chondrite-like NEOs are located on high-inclination orbits (mostly 12-35 degree), clumping near the Phocaea and Maria families; these orbits have relatively long life times and low probabilities of collisions with the Earth. Therefore, most H-chondrite-like NEOs observed via telescopic observations are unlikely to relate to H-chondrite meteorites found on Earth (Brož et al. 2023).

There are several additional and independent observables pointing at Massalia as being the main source of L chondrites. First, the size of (20) Massalia, with diameter 130-150 km (e.g. Alí-Lagoa et al. 2018, 2020, Herald et al. 2019, Mainzer et al. 2019) is within the estimated size range of the L-chondrite parent body (100–230 km; Haack et al. 1996; Gail & Trieloff 2019). Second, IRAS and COBE observations of the zodiacal cloud (Sykes 1990, Reach et al. 1997) revealed the existence of a toroidal dust band with exactly the same orbital inclination 1.4 degree as (20) Massalia (Nesvorný et al. 2003), indicating that this family is still an active source of meteorites hitting the Earth today. Remarkably, the observed Massalia family has a steep SFD, with a power-law index of −2.8. An extrapolation of this SFD down to 100-μm size, corresponding to the dominant size of cosmic dust particles (Love & Brownlee 1993), makes the SDF reach the measured dust abundance of the 1.4-degree band (Fig. 4). At one-meter size, the interpolated SFD implies a number of family members of 10 to 30 × $10^{10}$, and a number of escaped NEOs from the family of 30 to 80 × $10^8$, i.e. about four to six times the estimated number of meter-size NEOs from the LL-chondrite-like Flora family (Brož et al. 2023). It follows that the identification of Massalia as the source of L chondrites and of the IRAS 1.4-degree dust band naturally solves the so-called "asteroid-meteorite conundrum" (Vernazza et al. 2008, Binzel et al. 2015) in the case of L and LL-chondrite-like bodies. This conundrum stems from the apparent inconsistency between the compositional distribution of observed NEOs in the sky, and that of meteorites falls. Specifically, L-chondrite-like NEOs are 7 times less abundant than their LL-chondrite-like counterparts (Marsset et al. 2022), whereas L-chondrite meteorite falls are a factor 4.5 more abundant than LL-chondrite falls (Gattacceca et al. 2022). Here, this apparent contradiction is naturally explained by the much steeper SFD of Massalia compared to Flora (Brož et al. 2023; SI Fig. 5).

Next, we explored the collisional evolution of the Massalia family by means of numerical simulations (Appendix C). Assuming that the initial SFD was as steep as the observed SFD between 5 and 2.5 km, we obtained a best-fit age of (450 ± 50)

My (SI Fig. 7). This is highly compatible with the estimated age of the break up of the L-chondrite parent body (Heck et al. 2017, Schmitz et al. 2019, Liao et al. 2020) and the argon isotopic age of shocked L chondrites (Swindle et al. 2014). However, the SFD cannot remain steep for hundreds of My. In particular, the 1.4-degree dust band must have been formed relatively recently. According to our simulations, the formation of a second Massalia family about 40 My ago explains all observations. Consequently, it is a similar case as the Vesta family and the two basins present on its surface (Schenk et al. 2012). We repeated the same analysis for Gefion, the previously favored parent family of L chondrites, and confirmed that, with an estimated age of (1500 ± 200) My (Brož et al. 2023), it is too old compared to the breakup of the L-chondrite parent body, in agreement with previous works (Aljbaae et al. 2019). For Juno, which is also classified as L/LL, no dust band is observed, which indicates that its relatively steep SFD must bend and become shallow below the present-day observation completeness limit.

Our findings make several predictions. First, the Vera-C.-Rubin's LSST survey (Ivezić et al. 2019) should find that the SFD of Massalia remains steep, with power-law index close to −2.8, down to observational completeness of the survey (~500 m for main-belt asteroids; LSST Science Collaboration et al. 2009). Second, the fraction of L-chondrite-like bodies in the NEO population must increase at sizes smaller than ~100 meters, corresponding to the size allowing a substantial number of Massalia asteroids to be transported to the $v_6$ resonance by the Yarkovsky effect. Below this threshold, the ratio of L-to-LL-like NEOs should begin to match the proportions of meteorites' fall statistics. This is true because both the Massalia and Flora families have low-inclination orbits and, therefore, high impact probabilities. Third, high-resolution adaptive-optics images of (20) Massalia should reveal a Vesta-like impact basin or a reaccumulated fresh surface, corresponding to the second collisional event, which is at the origin of the dust band. Finally, camera networks designed for monitoring fireballs and organizing recovery campaigns of meteorites (Colas et al. 2020, Spurný et al. 2020) should continue to find that recovered L chondrites preferentially originate from the inner main belt and reach the Earth via the $v_6$ or 3:1 resonances on relatively low-inclination orbits (mostly <13 degree; SI Figs. 8, 9; Appendix D, Jenniskens et al. 2019).

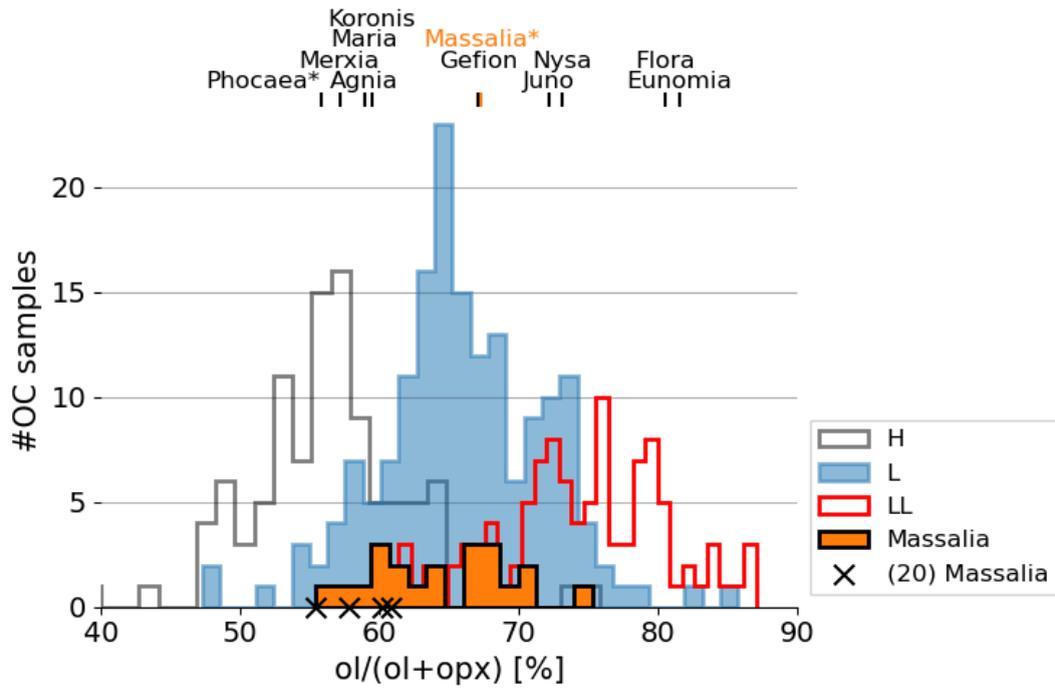

*Fig. 1: Measured mineralogy (in terms of olivine-to-pyroxene ratio; ol/(ol+opx)) of 15 asteroids from the Massalia family (retrieved from a total of 21 individual spectra), compared to the mineralogy of ordinary chondrites (H, L, LL types) from RELAB. Average values are indicated at the top for each S-type family (families marked with an asterisk had their largest member removed before computing the average). Measurements of (20) Massalia are indicated in the plot (X). While this body lays between the peaks of the H-chondrite and L-chondrite distributions, other family members are clearly compatible with L chondrites.*

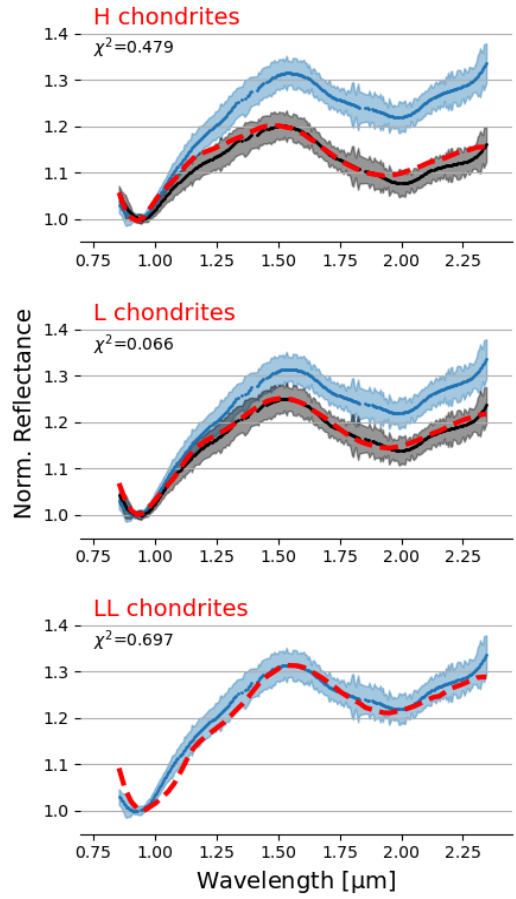

*Fig. 2: Comparison between the average reflectance spectra of ordinary H, L and LL chondrites (red) and the average spectrum of small Massalia family members, before (blue) and after (black) dereddening by use of the empirical function of Brunetto et al. (2006). The reflectance is normalized at the minimum 1-µm band value. Its uncertainty is typically 0.01; for the average spectrum we assign σ equal to the dispersion of spectra in the group. In the case of LL chondrites, the best spectral fit is obtained for $C_s=0$ (no dereddening; Appendix B) so we only show the original asteroid spectrum. The $\chi^2$ values of the fits between the dereddened asteroid spectrum and the OC spectra are indicated in each corresponding panel. It confirms the classification of the family as L-chondrite-like. See appendix B for additional information and for plots of other S-type families.*

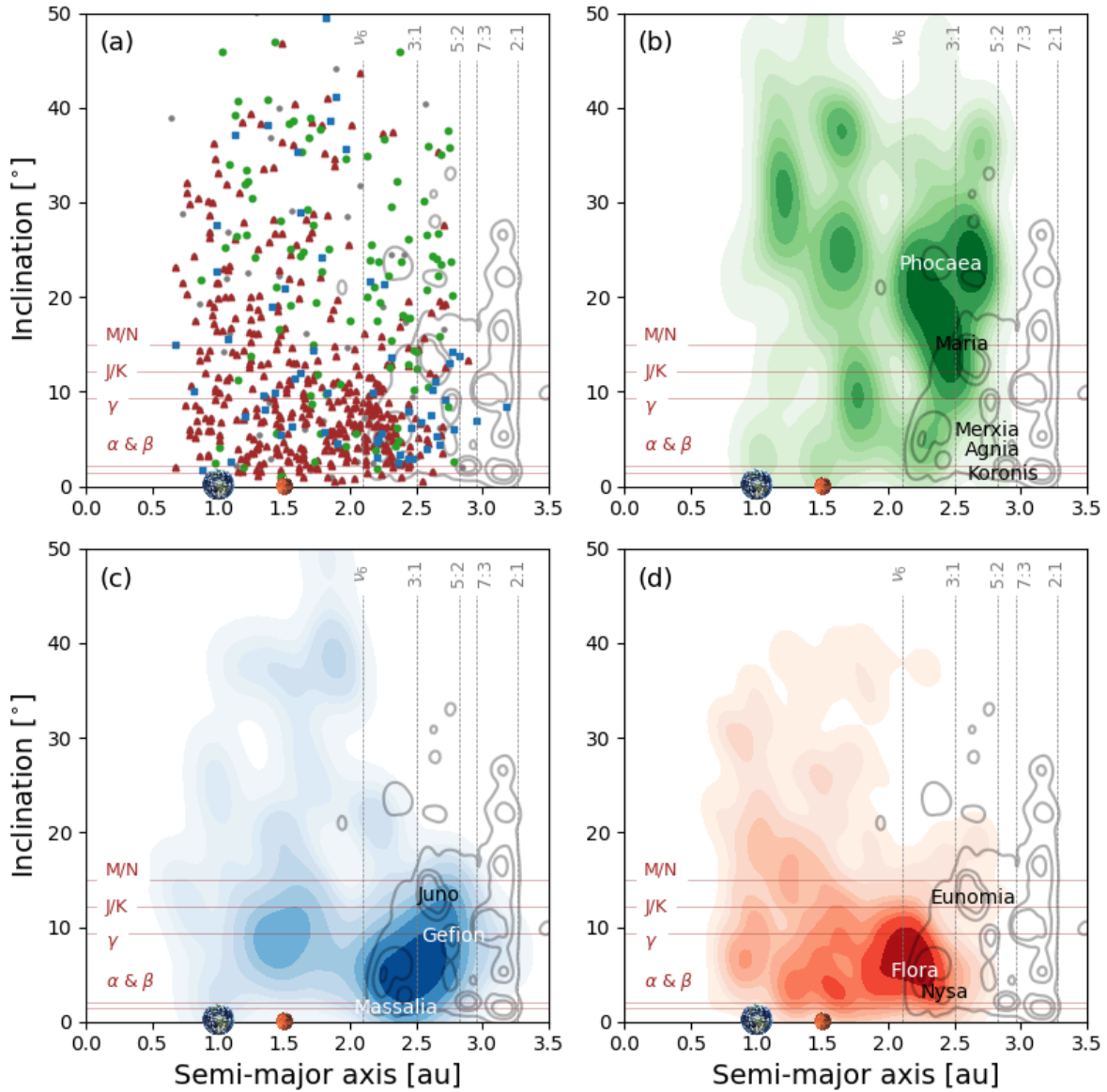

Fig. 3: (a) Orbital distribution of S-type NEOs, divided into classes of meteorite analogs: H (green circles), L (blue squares), LL (red triangles) and uncertain (gray circles). The gray contours correspond to the number density of objects in the asteroid belt. (b), (c) and (d): Density distribution of H-chondrite-like, L-chondrite-like and LL-chondrite-like NEOs, respectively. Selected S-type collisional families discussed in the text are labeled. L-chondrite-like NEOs are preferentially located around the 3:1 resonance, on low-inclination orbits near Massalia, whereas LL chondrite-like NEOs are close to Flora, and H chondrite-like NEOs are preferentially found on high-inclination orbits near Phocaea and Maria. The vertical lines indicate the location of major resonances in the asteroid belt. The horizontal lines indicate the model-derived inclination of some of the IRAS and COBE dust bands. With an orbital inclination of 1.4 degree, the α band intersects the Massalia family (Nesvorný et al. 2003), indicating that this family is an active source of extraterrestrial material presently being accreted by the Earth.

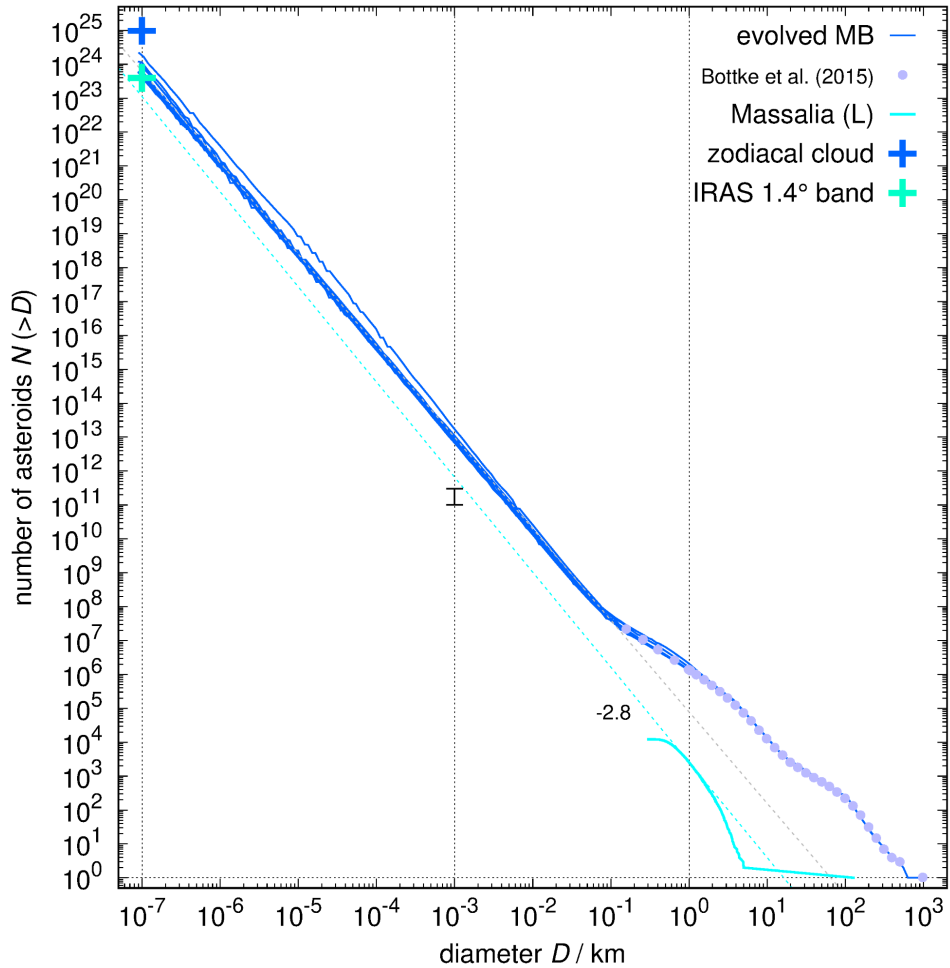

*Fig. 4: Observed size-frequency distribution of the Massalia family. Its extrapolation matches the dust abundance in the 1.4-degree IRAS dust band. The family contains 12172 asteroids, with the largest being 100-km-size (20) Massalia and the smallest approximately 0.5 km, as determined by the current observational limit (cyan). The power-law slope of the cumulative distribution is −2.8 (dotted). The number of dust particles forming the dust band is of the order of $4 \times 10^{23}$, assuming the dominant size of 100 µm (Love & Brownlee 1993). The expected number of L-chondrite-like meter-sized bodies is $10$ to $30 \times 10^{10}$ (error bar). For comparison, the distribution of the main belt is plotted, both observed (circles; Bottke et al. 2015) and extrapolated using a collisional model (blue).*

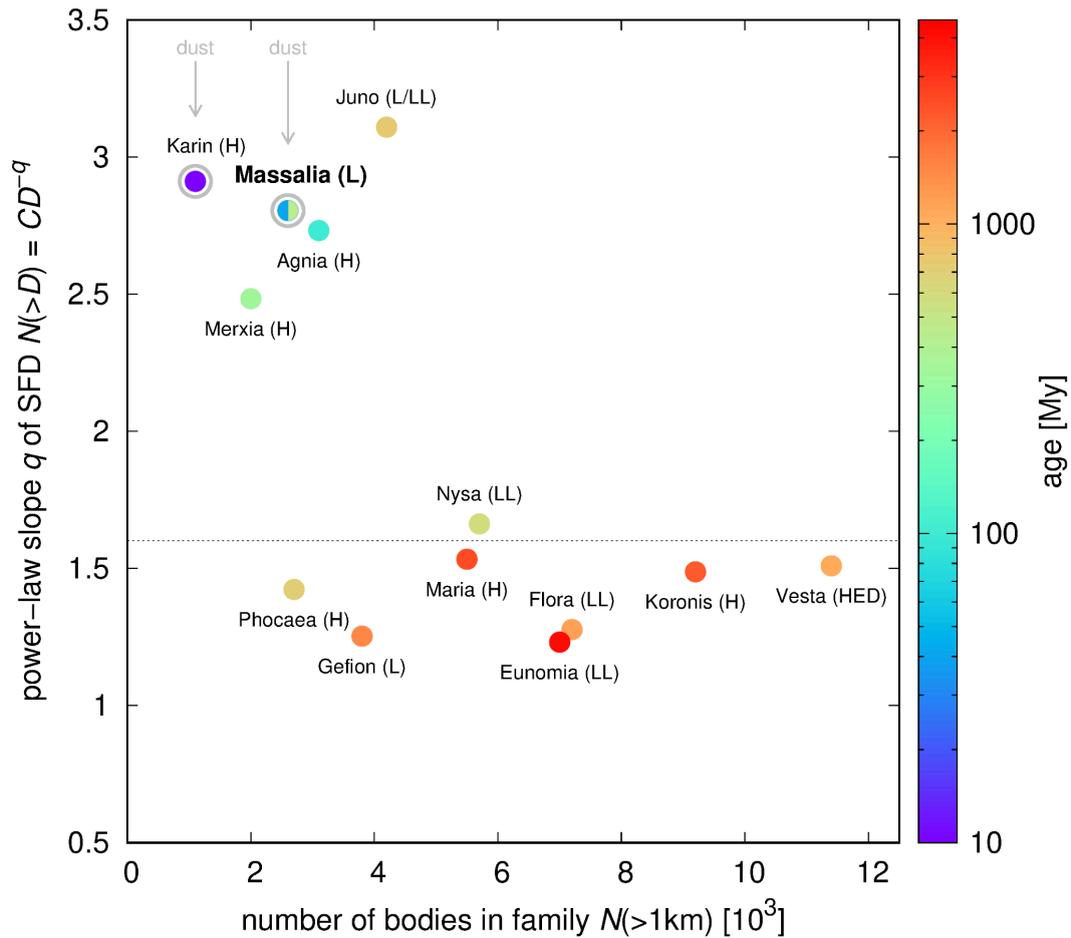

SI Fig. 5: Young S-type families must contribute more than old ones to the population of sub-km sized bodies. The observed numbers of >1 km bodies and the respective power-law slopes of the SFD above the observational completeness limit (from Brož et al. 2023) shows that young families exhibit steep slopes (close to −3), while old families have shallow slopes (close to −1.5; similarly as the main belt population). The age is indicated by color. The Massalia family is exceptional and, moreover, it is related to the 1.4-degree IRAS dust band. For the discussion of the Karin family, see Brož et al. (2023).

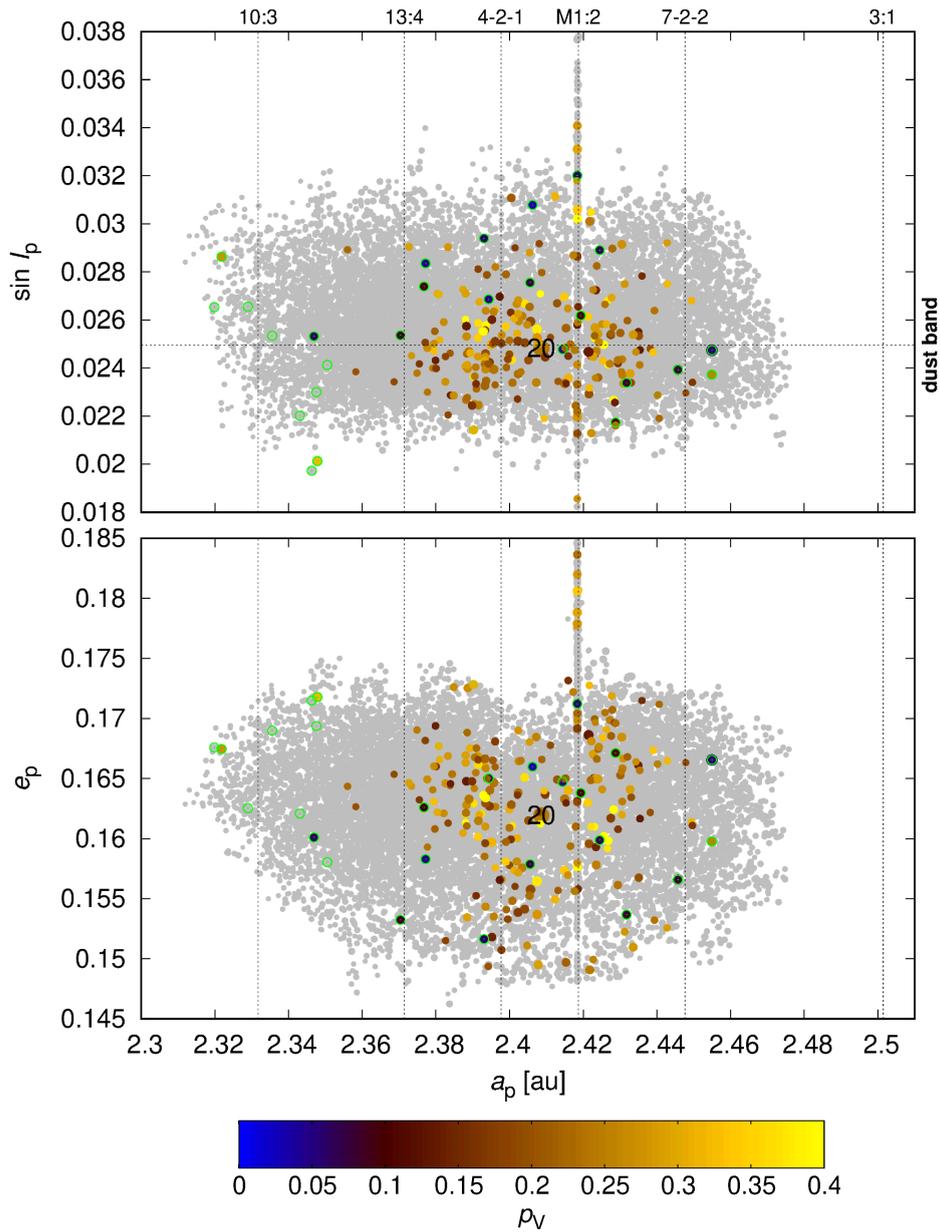

SI Fig. 6: Orbital elements of the Massalia family and of (20) Massalia match the 1.4-degree IRAS dust band. The proper semimajor axis versus the eccentricity (bottom) or the inclination (top) are plotted (Novaković and Radović 2019). Colors (blue to yellow) correspond to the geometric albedo (if known). The family contains 12172 asteroids, excluding a few interlopers (green). The locations of resonances are also indicated (vertical dotted lines), including the 3:1 mean-motion resonance with Jupiter and the 1:2 with Mars, which deliver bodies to the NEO space.

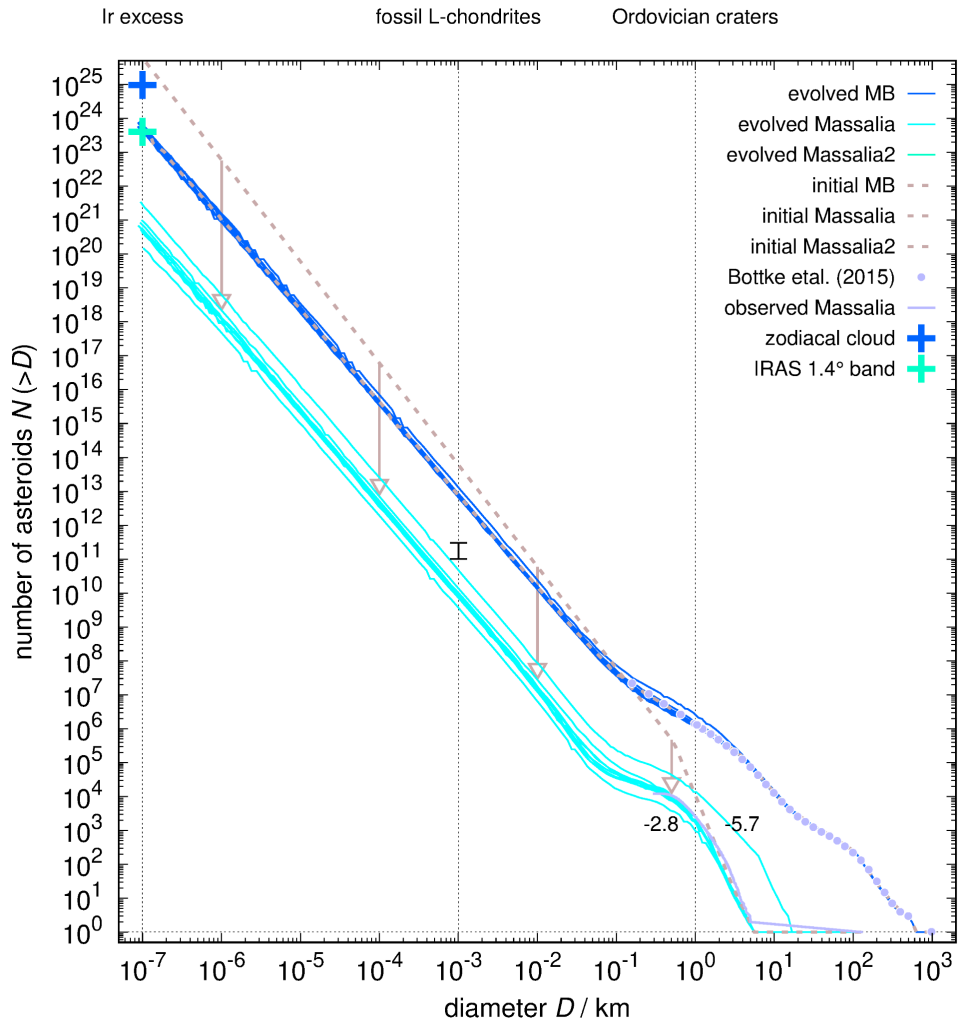

SI Fig. 7a: Synthetic size-frequency distribution of the Massalia family. The axes are the same as in Fig. 4. The initial conditions (brown) correspond to a steep slope (−5.7). The labels on top are the observations (Schmitz et al. 1997, Schmieder and Kring 2020) indicating a steep slope. After 470 My of collisional evolution, it is modified to a shallow slope (−2.8), i.e., very close to the observed value. The outcome depends on random collisions; at least 10 simulations must be computed to describe the statistical distribution. Nevertheless, the tail never matches the dust band, nor the expected number of L-chondrite-like bodies.

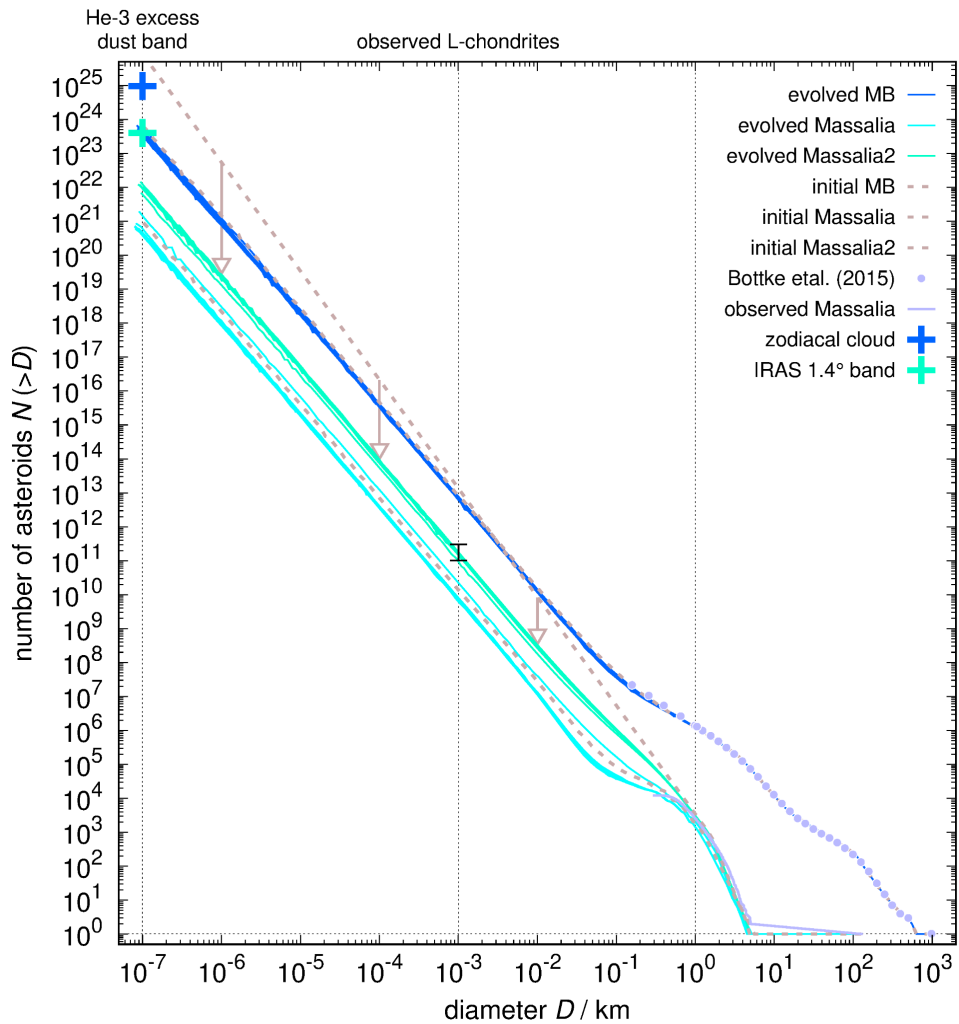

*SI Fig. 7b: Synthetic size-frequency distribution of the second Massalia family. The initial conditions correspond to a time of 430 My, when a second population forms with a steep slope (−5.7 and −3.1 below 1 km). The labels on top are the observations (Farley et al. 1998, Sykes 1990, Schmieder and Kring 2020) indicating a steep slope. After an additional 40 My of collisional evolution, it is modified to a shallow slope (−2.8). The resulting SFD of the first (cyan) and second (green) Massalia families matches the dust band and the number of L-chondrite-like bodies.*

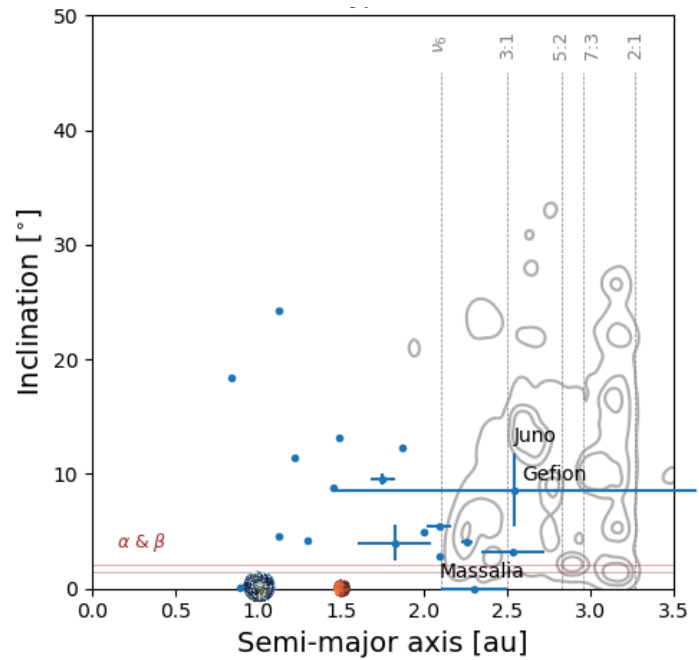

*SI Fig. 8: Pre-atmospheric orbital elements of 18 L-chondrite falls. The osculating semimajor axis versus the inclination are plotted with their uncertainties (error bars). Most L chondrites have inclinations lower than 10 deg, which points to the Massalia family as a source region with very low inclination. In particular, neither of the other two L-chondrite-like families, Gefion and Juno, have low inclinations. The axis ranges are as in Fig. 3. Data from Meier (2023); https://www.meteoriteorbits.info/.*

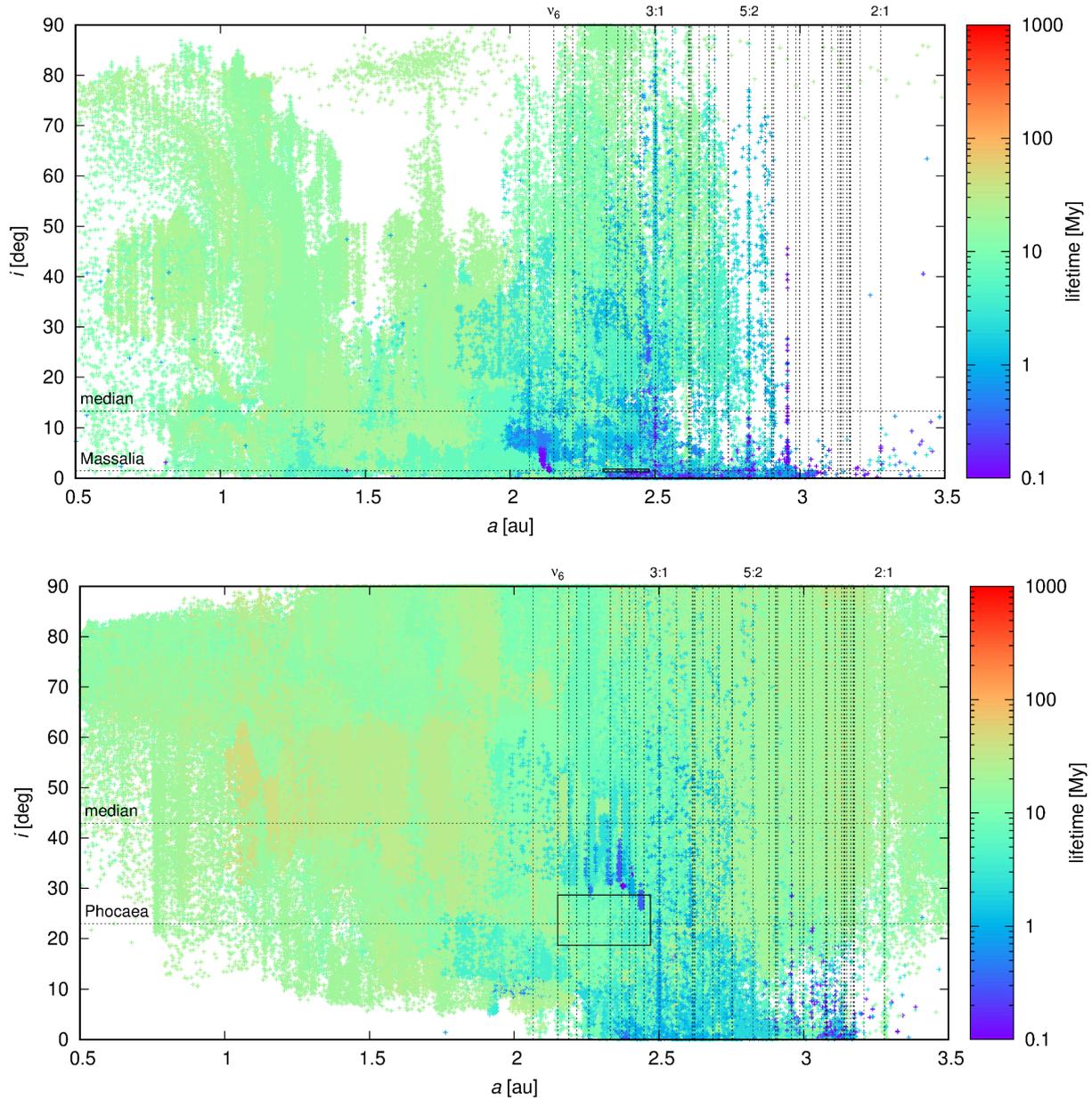

SI Fig. 9: Orbital evolution of meter-sized meteoroids originating from the Massalia family (top) and the Phocaea family (bottom). The mean semimajor axis versus the inclination is plotted, with colors corresponding to the overall life time of bodies in the NEO space. Our simulations included more than $10^3$ bodies, influenced by the Yarkovsky drift, gravitational resonances, as well as close encounters with planets. The original inclinations of the families (1.4 and 23 degrees, respectively; black box) are partially preserved also in the NEO space. In particular, it is difficult to scatter bodies to inclinations lower than their original inclination in the main-belt.

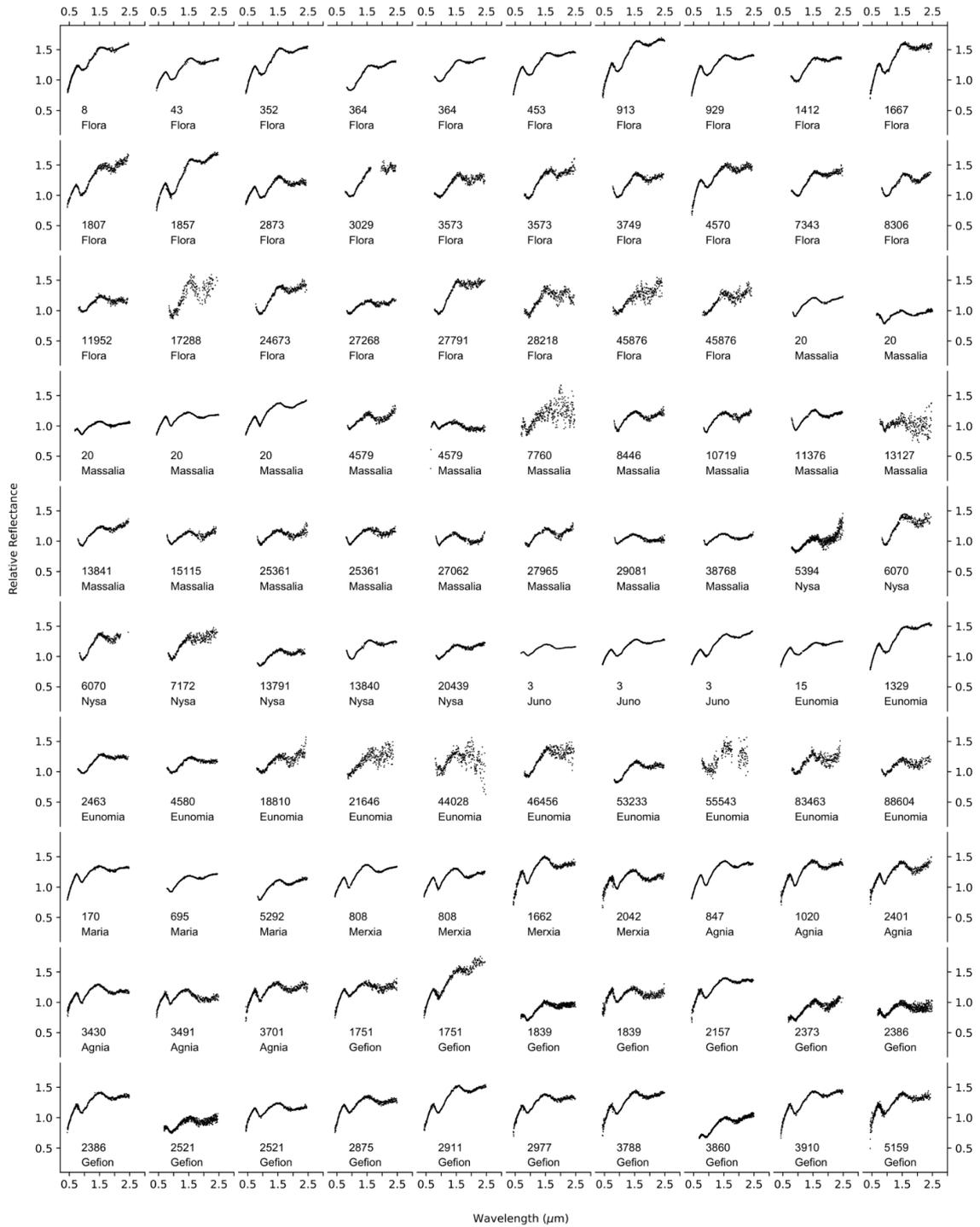

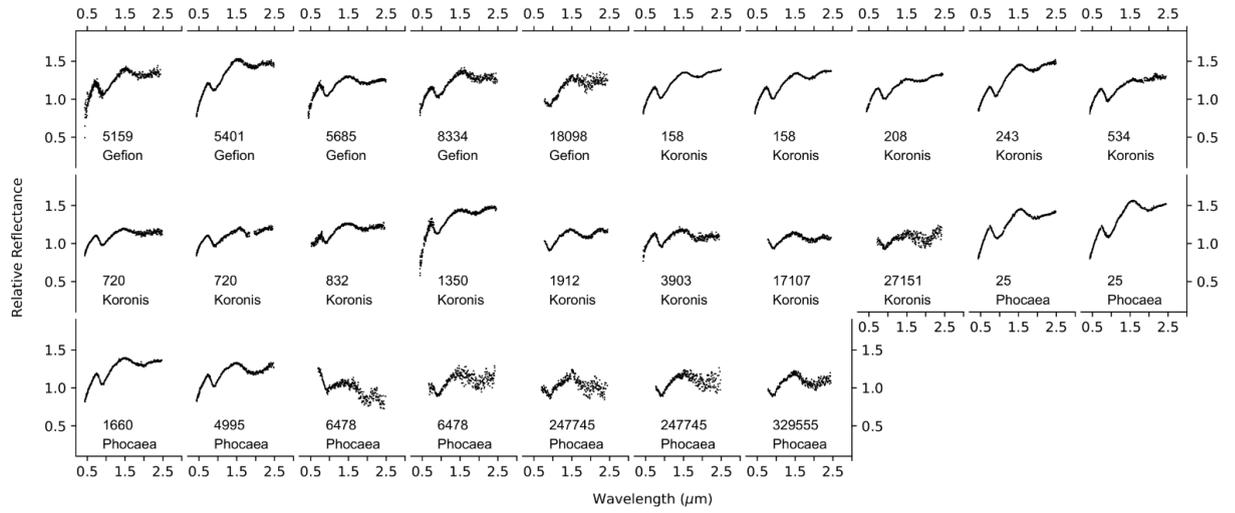

SI Fig. 10: Spectral data of S-type family members considered in this work. The number and family of each asteroid are indicated below the corresponding spectrum.

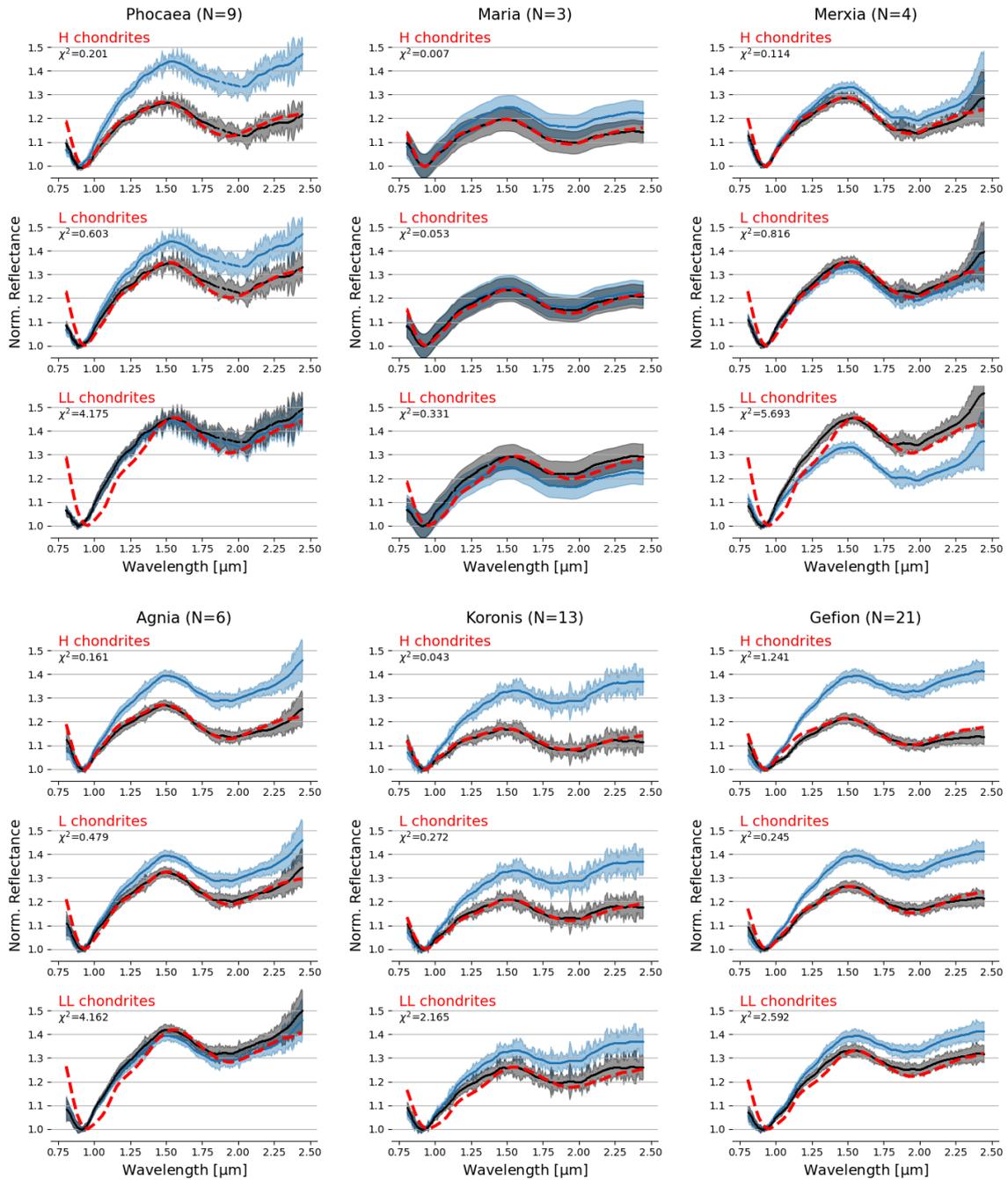

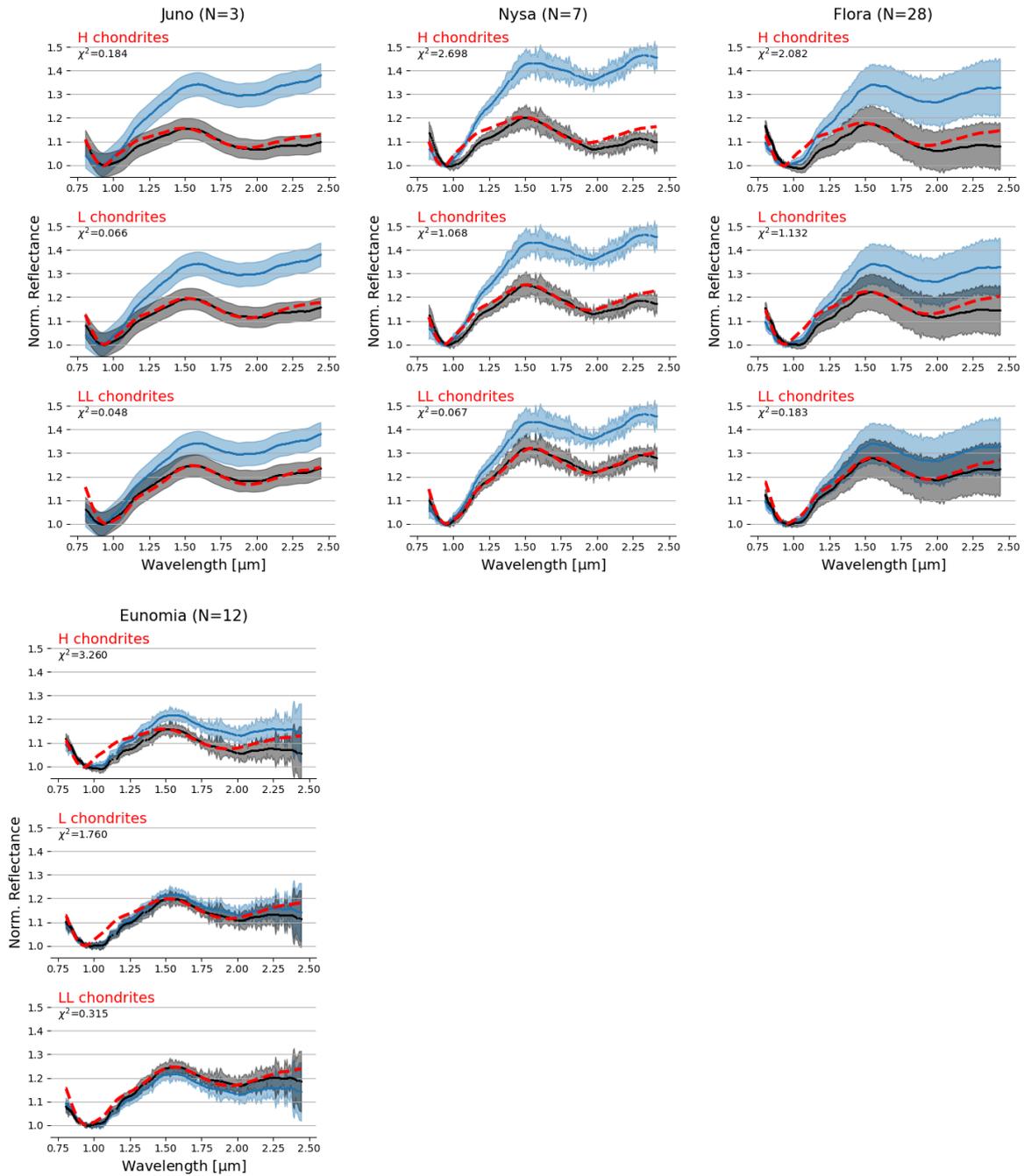

SI Fig. 11: Average family spectra compared to the average spectra of H, L and LL chondrites from RELAB, with grain sizes in the 0–45 µm range. The reduced chi square of the spectral fits is indicated in each panel. See Appendix B for more information about the method.

| Family | Class | ol/(ol+opx) [%] | $\chi^2$ | | |
|---|---|---|---|---|---|
| | | | H | L | LL |
| **Phocaea** | **H** | **55.78** | **0.201** | **0.603** | **4.175** |
| Merxia | H | 57.13 | 0.114 | 0.816 | 5.693 |
| Agnia | H | 58.92 | 0.161 | 0.479 | 4.162 |
| Maria | H | 58.95 | 0.007 | 0.053 | 0.331 |
| Koronis | H | 59.42 | 0.043 | 0.272 | 2.165 |
| Gefion | L | 67.10 | 1.241 | 0.245 | 2.592 |
| **Massalia** | **L** | **67.32** | **0.479** | **0.066** | **0.697** |
| Juno | L/LL | 72.21 | 0.184 | 0.066 | 0.048 |
| **Nysa** | **LL** | **73.11** | **2.698** | **1.068** | **0.067** |
| Flora | LL | 80.48 | 2.082 | 1.132 | 0.183 |
| Eunomia | LL | 81.51 | 3.260 | 1.760 | 0.315 |

*SI Table 1: Classification and measured mineralogy of major S-type families in the main belt. The last three columns indicate the measured $\chi^2$ between the average family spectra and each OC type (H, L or LL). The red color indicates the two families, Phocaea and Massalia, for which we find a differing OC classification with respect to previous works. The Nysa family (green) is characterized for the first time. The grey color indicates the best $\chi^2$ of each family. In the case of Juno, the close $\chi^2$ values do not allow to differentiate between mineralogical classes. The ol/(ol+opx) ratio and the shape of Juno's 1-µm band (SI Fig. 11) also make it intermediate between L and LL chondrites. Additional observations of smaller family members are needed to confirm the mineralogical group of the Juno family.*

# References

[1] Heck, P., Schmitz, B., Bottke, W. *et al.* Rare meteorites common in the Ordovician period. *Nature Astronomy*, vol. 1, 0035 (2017). doi:10.1038/s41550-016-0035

[2] Schmieder, M., and Kring, D. A., "Earth's Impact Events Through Geologic Time: A List of Recommended Ages for Terrestrial Impact Structures and Deposits", *Astrobiology*, vol. 20, no. 1, pp. 91–141, 2020. doi:10.1089/ast.2019.2085.

[3] Kenkmann, T., "The terrestrial impact crater record: A statistical analysis of morphologies, structures, ages, lithologies, and more", *Meteoritics and Planetary Science*, vol. 56, no. 5, pp. 1024–1070, 2021. doi:10.1111/maps.13657.

[4] Schmitz, B., Farley, K. A., Goderis, S. *et al.* "An extraterrestrial trigger for the mid-Ordovician ice age: Dust from the breakup of the L-chondrite parent body", *Science Advances*, vol. 5, no. 9, p. eaax4184, 2019. doi:10.1126/sciadv.aax4184.

[5] Swindle, T. D., Kring, D. A., and Weirich, J. R., "$^{40}$Ar/$^{39}$Ar ages of impacts involving ordinary chondrite meteorites", *Geological Society of London Special Publications*, vol. 378, no. 1, pp. 333–347, 2014. doi:10.1144/SP378.6.

[6] Heymann, D., "On the Origin of Hypersthene Chondrites: Ages and Shock Effects of Black Chondrites", *Icarus*, vol. 6, no. 1, pp. 189–221, 1967. doi:10.1016/0019-1035(67)90017-6.

[7] Marti, K. and Graf, T., "Cosmic-Ray Exposure History of Ordinary Chondrites", *Annual Review of Earth and Planetary Sciences*, vol. 20, p. 221, 1992. doi:10.1146/annurev.ea.20.050192.001253.

[8] Rubin, A. E., "Metallic Copper in Ordinary Chondrites", *Meteoritics*, vol. 29, no. 1, p. 93, 1994. doi:10.1111/j.1945-5100.1994.tb00659.x.

[9] Bischoff, A., Schleiting, M., and Patzek, M., "Shock stage distribution of 2280 ordinary chondrites—Can bulk chondrites with a shock stage of S6 exist as individual rocks?", *Meteoritics and Planetary Science*, vol. 54, no. 10, pp. 2189–2202, 2019. doi:10.1111/maps.13208.

[10] Korochantseva, E. V., Trieloff, M., Lorenz, C. A. *et al.* "L-chondrite asteroid breakup tied to Ordovician meteorite shower by multiple isochron $^{40}$Ar-$^{39}$Ar dating", *Meteoritics and Planetary Science*, vol. 42, no. 1, pp. 113–130, 2007. doi:10.1111/j.1945-5100.2007.tb00221.x.

[11] Haack, H., Farinella, P., Scott, E. R. D., and Keil, K., "Meteoritic, Asteroidal, and Theoretical Constraints on the 500 MA Disruption of the L Chondrite Parent Body", *Icarus*, vol. 119, no. 1, pp. 182–191, 1996. doi:10.1006/icar.1996.0010.

[12] Greenwood, R. C., Burbine, T. H., and Franchi, I. A., "Linking asteroids and meteorites to the primordial planetesimal population", *Geochimica et Cosmochimica Acta*, vol. 277, pp. 377–406, 2020. doi:10.1016/j.gca.2020.02.004.

[13] Schmitz, B., Peucker-Ehrenbrink, B., Lindström, M., and Tassinari, M., "Accretion rates of meteorites and cosmic dust in the Early Ordovician.", *Science*, vol. 278, no. 5335, pp. 88–90, 1997. doi:10.1126/science.278.5335.88.

[14] Schmitz, B., Tassinari, M., and Peucker-Ehrenbrink, B., "A rain of ordinary chondritic meteorites in the early Ordovician", *Earth and Planetary Science Letters*, vol. 194, no. 1–2, pp. 1–15, 2001. doi:10.1016/S0012-821X(01)00559-3.

[15] Terfelt, F. and Schmitz, B., "Asteroid break-ups and meteorite delivery to Earth the past 500 million years", *Proceedings of the National Academy of Science*, vol. 118, no. 24, 2021. doi:10.1073/pnas.2020977118.

[16] Nesvorný, D., Brož, M., and Carruba, V., "Identification and Dynamical Properties of Asteroid Families", in *Asteroids IV*, P. Michel, F. E. DeMeo, and W. F. Bottke (eds.), University of Arizona Press, Tucson, pp. 297–321, 2015. doi:10.2458/azu_uapress_9780816532131-ch016.

[17] Gaffey, M. J., Bell, J. F., Brown, H. R., Burbine, T. H., Piatek, J. L., Reed, K. L. and Chaky, D. A. "Mineralogical Variations within the S-Type Asteroid Class", *Icarus*, vol. 106, no. 2, pp. 573–602, 1993. doi:10.1006/icar.1993.1194.

[18] Nakamura, T., Noguchi, T., Tanaka, M. *et al.* "Itokawa Dust Particles: A Direct Link Between S-Type Asteroids and Ordinary Chondrites", *Science*, vol. 333, no. 6046, p. 1113, 2011. doi:10.1126/science.1207758.

[19] Vernazza, P., Zanda, B., Binzel, R. P. *et al.* "Multiple and Fast: The Accretion of Ordinary Chondrite Parent Bodies", *Astrophysical Journal*, vol. 791, no. 2, 2014. doi:10.1088/0004-637X/791/2/120.

[20] Aljbaae, S., Souchay, J., Prado, A. F. B. A., and Chanut, T. G. G., "A dynamical study of the Gefion asteroid family", *Astronomy and Astrophysics*, vol. 622, 2019. doi:10.1051/0004-6361/201834470.

[21] Pieters, C. M. and Hiroi, T., "RELAB (Reflectance Experiment Laboratory): A NASA Multiuser Spectroscopy Facility", p. 1720, 2004.

[22] Milliken, R. E., Hiroi, T., and Patterson, W., "The NASA Reflectance Experiment Laboratory (RELAB) Facility: Past, Present, and Future", no. 1903, p. 2058, 2016.

[23] Brunetto, R., Vernazza, P., Marchi, S., Birlan, M., Fulchignoni, M., Orofino, V., Strazzulla, G., "Modeling asteroid surfaces from observations and irradiation experiments: The case of 832 Karin", *Icarus*, vol. 184, no. 2, pp. 327–337, 2006. doi:10.1016/j.icarus.2006.05.019.

[24] Shkuratov, Y., Starukhina, L., Hoffmann, H., and Arnold, G., "A Model of Spectral Albedo of Particulate Surfaces: Implications for Optical Properties of the Moon", *Icarus*, vol. 137, no. 2, pp. 235–246, 1999. doi:10.1006/icar.1998.6035.

[25] Binzel, R. P., DeMeo, F. E., Turtelboom, E. V. *et al.* "Compositional distributions and evolutionary processes for the near-Earth object population: Results from the MIT-Hawaii Near-Earth Object Spectroscopic Survey (MITHNEOS)", *Icarus*, vol. 324, pp. 41–76, 2019. doi:10.1016/j.icarus.2018.12.035.

[26] Granvik, M., Morbidelli, A., Jedicke, R. *et al.* "Debiased orbit and absolute-magnitude distributions for near-Earth objects", *Icarus*, vol. 312, pp. 181–207, 2018. doi:10.1016/j.icarus.2018.04.018.

[27] Nesvorný, D., Bottke, W. F., Levison, H. F., and Dones, L., "Recent Origin of the Solar System Dust Bands", *Astrophysical Journal*, vol. 591, no. 1, pp. 486–497, 2003. doi:10.1086/374807.

[28] Vokrouhlický, D., Brož, M., Bottke, W. F., Nesvorný, D., and Morbidelli, A., "Yarkovsky/YORP chronology of asteroid families", *Icarus*, vol. 182, no. 1, pp. 118–142, 2006. doi:10.1016/j.icarus.2005.12.010.

[29] Spoto, F., Milani, A., and Knežević, Z., "Asteroid family ages", *Icarus*, vol. 257, pp. 275–289, 2015. doi:10.1016/j.icarus.2015.04.041.

[30] Brož, M., Marsset, M., Vernazza, P., DeMeo, F., Binzel, R., Vokrouhlický, D., Nesvorný, D., "Contribution of S-type asteroid families to NEOs and meteoroid populations", Submitted to *Astronomy and Astrophysics*.

[31] Marsset, M., DeMeo, F. E., Burt, B. *et al.* "The Debiased Compositional Distribution of MITHNEOS: Global Match between the Near-Earth and Main-belt Asteroid Populations, and Excess of D-type Near-Earth Objects", *Astronomical Journal*, vol. 163, no. 4, 2022. doi:10.3847/1538-3881/ac532f.

[32] Kozai, Y., "Secular perturbations of asteroids with high inclination and eccentricity", *Astronomical Journal*, vol. 67, pp. 591–598, 1962. doi:10.1086/108790.

[33] Thomas, C. A. and Binzel, R. P., "Identifying meteorite source regions through near-Earth object spectroscopy", *Icarus*, vol. 205, no. 2, pp. 419–429, 2010. doi:10.1016/j.icarus.2009.08.008.

[34] de León, J., Licandro, J., Serra-Ricart, M., Pinilla-Alonso, N., and Campins, H., "Observations, compositional, and physical characterization of near-Earth and Mars-crosser asteroids from a spectroscopic survey", *Astronomy and Astrophysics*, vol. 517, 2010. doi:10.1051/0004-6361/200913852.

[35] Dunn, T. L., Burbine, T. H., Bottke, W. F., and Clark, J. P., "Mineralogies and source regions of near-Earth asteroids", *Icarus*, vol. 222, no. 1, pp. 273–282, 2013. doi:10.1016/j.icarus.2012.11.007.

[36] Alí-Lagoa, V., Müller, T. G., Usui, F., and Hasegawa, S., "The AKARI IRC asteroid flux catalogue: updated diameters and albedos", *Astronomy and Astrophysics*, vol. 612, 2018. doi:10.1051/0004-6361/201731806.

[37] Alí-Lagoa, V., Müller, T. G., Kiss, C., *et al.*, "Thermal properties of large main-belt asteroids observed by Herschel PACS", *Astronomy and Astrophysics*, vol. 638, 2020. doi:10.1051/0004-6361/202037718.

[38] Herald, D., Frappa, E., Gault, D., *et al.*, "Small Bodies Occultations Bundle V3.0", *NASA Planetary Data System*, p. 3, 2019. doi:10.26033/ap0g-wf63.

[39] Mainzer, A. K., Bauer, J. M., Cutri, R. M. *et al.* "NEOWISE Diameters and Albedos V2.0", NASA Planetary Data System, 2019. doi:10.26033/18S3-2Z54.

[40] Gail, H.-P. and Trieloff, M., "Thermal history modelling of the L chondrite parent body", *Astronomy and Astrophysics*, vol. 628, 2019. doi:10.1051/0004-6361/201936020.

[41] Sykes, M. V., "Zodiacal dust bands: Their relation to asteroid families", *Icarus*, vol. 85, no. 2, pp. 267–289, 1990. doi:10.1016/0019-1035(90)90117-R.

[42] Reach, W. T., Franz, B. A., and Weiland, J. L., "The Three-Dimensional Structure of the Zodiacal Dust Bands", *Icarus*, vol. 127, no. 2, pp. 461–484, 1997. doi:10.1006/icar.1997.5704.

[43] Love, S. G. and Brownlee, D. E., "A Direct Measurement of the Terrestrial Mass Accretion Rate of Cosmic Dust", *Science*, vol. 262, no. 5133, pp. 550–553, 1993. doi:10.1126/science.262.5133.550.

[44] Vernazza, P., Binzel, R. P., Thomas, C. A., DeMeo, F. E., Bus, S. J., Rivkin, A. S., and Tokunaga, A. T., "Compositional differences between meteorites and near-Earth asteroids", *Nature*, vol. 454, no. 7206, pp. 858–860, 2008. doi:10.1038/nature07154.

[45] Gattacceca, J., MCCubbin, F. M., Grossman, J., Bouvier, A., Chabot, N. L., D'Orazio, M., Goodrich, C., *et al.*, "The Meteoritical Bulletin, No. 110", *Meteoritics and Planetary Science*, vol. 57, no. 11, pp. 2102–2105, 2022. doi:10.1111/maps.13918.

[46] Liao, S., Huyskens, M. H., Yin, Q.-Z., and Schmitz, B., "Absolute dating of the L-chondrite parent body breakup with high-precision U-Pb zircon geochronology from Ordovician limestone", *Earth and Planetary Science Letters*, vol. 547, 2020. doi:10.1016/j.epsl.2020.116442.

[47] Ivezić, Ž., Kahn, S. M., Tyson, J. A. *et al.* "LSST: From Science Drivers to Reference Design and Anticipated Data Products", *Astrophysical Journal*, vol. 873, no. 2, 2019. doi:10.3847/1538-4357/ab042c.

[48] LSST Science Collaboration, "LSST Science Book, Version 2.0", arXiv e-prints, 2009. doi:10.48550/arXiv.0912.0201.

[49] Schenk, P., O'Brien, D. P., Marchi, S., Gaskell, R., Preusker, F., Roatsch, T., Jaumann, R., *et al.*, "The Geologically Recent Giant Impact Basins at Vesta's South Pole", *Science*, vol. 336, no. 6082, pp. 694–697, 2012. doi:10.1126/science.1223272.

[50] Colas, F., Zanda, B., Bouley, S. *et al.* "FRIPON: a worldwide network to track incoming meteoroids", *Astronomy and Astrophysics*, vol. 644, 2020. doi:10.1051/0004-6361/202038649.

[51] Spurný, P., Borovička, J., Shrbený, L., "The Žďár nad Sázavou meteorite fall: Fireball trajectory, photometry, dynamics, fragmentation, orbit, and meteorite recovery", *Meteoritics and Planetary Science*, vol. 55, no. 2, pp. 376–401, 2020. doi:10.1111/maps.13444.

[52] Jenniskens, P., Utas, J. Yin, Q.-Z., *et al.* "The Creston, California, meteorite fall and the origin of L chondrites", *Meteoritics and Planetary Science*, vol. 54, no. 4, pp. 699–720, 2019. doi:10.1111/maps.13235.

[53] Bottke, W. F., Brož, M., O'Brien, D. P., Campo Bagatin, A., Morbidelli, A., Marchi, S., "The Collisional Evolution of the Main Asteroid Belt", in *Asteroids IV*, P. Michel, F. E. DeMeo, and W. F. Bottke (eds.), University of Arizona Press, Tucson, pp. 701–724, 2015. doi:10.2458/azu_uapress_9780816532131-ch036.

*Acknowledgements*

We are thankful to Josh Emery, Maggie McAdam and Lauren McGraw who have been involved in the spectral survey that obtained one of the Massalia spectrum. Observations reported here were obtained at the NASA Infrared Telescope Facility, which is operated by the University of Hawaii under contract 80HQTR19D0030 with the National Aeronautics and Space Administration. The authors acknowledge the sacred nature of Maunakea and appreciate the opportunity to observe from the mountain. The MIT component of this work is supported by NASA grant 80NSSC18K0849. Any opinions, findings, and conclusions or recommendations expressed in this article are those of the authors and do not necessarily reflect the views of the National Aeronautics and Space Administration. This work has been supported by the Czech Science Foundation through grant 21-11058S (M. Brož).

*Author contributions*

M. Marsset and P. Vernazza led the spectral analysis. M. Brož computed collisional and orbital simulations. The three first authors jointly wrote the manuscript. C. Thomas provided most observations of small Massalia family members. F. DeMeo and R. Binzel were co-PIs of the MITHNEOS when the presented NEO observations were performed. V. Reddy, A. McGraw, C. Avdellidou and B. Carry provided additional spectroscopic observations presented in this work. B. Burt, S. Slivan and D. Polishook provide continuous support for MITHNEOS. All authors contributed inputs for the writing of the manuscript.

*Competing interests*

The authors declare no competing interests.

Appendix A: IRTF observations and data reduction

Most reflectance spectra of Massalia asteroids presented in this work were acquired by co-I C. Thomas using the SpeX spectrograph (Rayner et al. 2003) on NASA's 3-meter Infrared Telescope Facility (IRTF, Mauna Kea; Hawaii) to collect the data over the 0.8 to 2.5 μm wavelength range. Data reduction and spectral extraction followed the procedure outlined in Binzel et al. (2019). We summarize it briefly here. Reduction of the spectral images was performed with the Image Reduction and Analysis Facility (IRAF) and Interactive Data Language (IDL), using the Autospex software tool to automatically write sets of command files (Rivkin et al. 2005). Reduction steps for the science targets and their corresponding calibration stars included trimming the images, creating a bad pixel map, flat-fielding the images, sky subtracting between AB image pairs, tracing the spectra in both the wavelength and spatial dimensions, co-adding the spectral images, extracting the spectra, performing wavelength calibration, and correcting for air-mass differences between the asteroids and the corresponding solar analogs. Finally, the resulting asteroid spectra were divided by the mean stellar spectra to remove the solar gradient.

Additional spectra were compiled from the SMASS database (Bus & Binzel 2002, Burbine & Binzel 2002, http://smass.mit.edu/catalog.php) for Massalia and 10 more S-type families with more than 1,000 members (as defined in Nesvorný et al. 2015): Flora, Massalia, Nysa and Phocaea for the inner belt (inward of the 3:1 mean-motion resonance with Jupiter), Eunomia, Juno, Maria, Gefion, Merxia and Agnia for the middle belt (between 3:1 and 5:2), and Koronis for the outer belt (beyond 5:2). Additional spectra were provided by co-Is P. Vernazza, C. Avdellidou, B. Carry, V. Reddy and A. McGraw. Thumbnail images of the spectral data of family members are shown in SI Fig. 10. Observing circumstances for the new observations are provided in SI Table 2.

NEO spectra analyzed in this work are previously published data from Binzel et al. (2019) and Marsset et al. (2022), acquired as part of the MIT-Hawaii Near-Earth Object Spectroscopic Survey (MITHNEOS). This survey obtained NIR (0.8–2.5 μm) spectroscopic measurements for around one thousand different NEOs with SpeX. From this dataset, we identified 621 unique S-type NEOs. Details about the observing strategy and reduction method of MITHNEOS can be found in the aforementioned papers.

| Asteroid | Date | Starting Time | Exposure Time [s] | Airmass | V mag. | Δ [au] | r [au] | α [deg.] |
|---|---|---|---|---|---|---|---|---|
| 20 | 2017 12 27 | 09:45 | 44 | 1.00 | 8.9 | 1.102 | 2.073 | 5.8 |
| 20 | 2014 12 22 | 15:42 | 600 | 1.31 | 11.4 | 2.434 | 2.275 | 23.8 |
| 182 | 2017 01 20 | 13:31 | 480 | 1.0 | 12.7 | 1.466 | 2.31 | 15.8 |
| 4579 | 2015 03 17 | 05:13 | 12240 | 1.09 | 18.3 | 1.792 | 2.485 | 19.5 |
| 8446 | 2015 12 05 | 13:25 | 6960 | 1.18 | 17.7 | 1.272 | 2.243 | 5.7 |
| 10719 | 2015 10 13 | 14:36 | 2640 | 1.11 | 17.3 | 1.192 | 2.057 | 18.2 |
| 11376 | 2016 03 10 | 09:43 | 10320 | 1.33 | 17.6 | 1.201 | 2.151 | 10.4 |
| 13841 | 2015 02 26 | 11:10 | 3600 | 1.03 | 16.7 | 1.096 | 2.086 | 0.9 |
| 15115 | 2016 09 19 | 10:40 | 8160 | 1.08 | 17.7 | 1.626 | 2.631 | 0.6 |
| 25361 | 2016 03 09 | 08:24 | 6960 | 1.13 | 17.6 | 1.219 | 2.207 | 3.4 |
| 25361 | 2016 03 10 | 08:10 | 3840 | 1.15 | 17.7 | 1.222 | 2.209 | 3.9 |
| 27062 | 2015 03 17 | 12:21 | 5520 | 1.23 | 16.9 | 1.147 | 2.138 | 3.1 |
| 27965 | 2016 11 11 | 13:28 | 3840 | 1.36 | 16.9 | 1.058 | 2.046 | 2.7 |
| 29081 | 2016 10 14 | 11:10 | 2880 | 1.01 | 17.2 | 0.987 | 1.962 | 8.5 |
| 38768 | 2015 10 13 | 09:03 | 6240 | 1.07 | 16.9 | 0.982 | 1.979 | 1.8 |
| 4579 | 2023 04 22 | 09:28 | 1920 | 1.15 | 16.1 | 1.098 | 2.101 | 2.5 |
| 7760 | 2023 04 22 | 12:39 | 1920 | 1.32 | 17.2 | 1.518 | 2.461 | 10.3 |
| 13127 | 2023 04 22 | 10:19 | 1920 | 1.20 | 17.2 | 1.076 | 2.073 | 4.9 |
| 5394 | 2023 06 12 | 06:10 | 1080 | 1.33 | 17.6 | 1.611 | 2.468 | 15.7 |

*SI Table 2: Observing conditions of the new spectra presented in this work. For each observation, we provide the asteroid number, the observing date and time, the total exposure time in seconds, as well as the air mass, visible magnitude (V mag.), geocentric distance (Δ), heliocentric distance (r) and phase angle of the asteroid during the observations.*

Appendix B: Mineralogy of asteroid families

The mineralogy of S-type asteroid families was investigated through direct spectral comparison with spectra of ordinary chondrites downloaded from the Reflectance Experiment Laboratory database (RELAB), the largest publicly available dataset of meteorites (Pieters & Hiroi 2004; Milliken et al. 2016). We retrieved a total of 957 visible and near-infrared spectra of OC meteorites, rejected spectra of meteorites labeled as dark or black ordinary chondrites (Britt & Pieters 1991; Reddy et al. 2014; Kohout et al. 2014, 2020; DeMeo et al. 2022), and considered the remaining subsample of spectra obtained for grain sizes within the 0–45 µm range, which usually provide the best spectral analogs for S-type asteroids (Vernazza et al. 2014).

S-type asteroids and OC meteorites are characterized by strong 1 and 2-µm silicate absorption bands in their visible and near-infrared reflectance spectra that are diagnostic of surface mineralogy (e.g., Cloutis et al. 1986; Gaffey et al. 1993). We performed a mineralogical analysis of S-type asteroids and ordinary chondrites using Vernazza et al. (2008)'s implementation of the Shkuratov et al. (1999) radiative transfer model, which allows simulating light interaction with a mixed material to produce an observed reflectance spectrum. Our spectra were modeled following the procedure described by Vernazza et al. (2008) and Binzel et al. (2019), using the meteorite sample of Dunn et al. (2010) for model calibration. The end-members minerals consisted of the most commonly found minerals in ordinary chondrites: olivine and orthopyroxene with a 70%/30% Fe/Mg mineralogy, and chromite. The resulting spectra were reddened by use of the empirical function of Brunetto et al. (2006), using their space weathering factor "$C_s$" as a free parameter in our solution. The best-fit model was then derived through a simple least-squares algorithm to varying fractions of olivine (ol) and orthopyroxene (opx) mineralogies, where these fractions are iteratively adjusted to find the resulting model spectrum having the minimum residuals with respect to the input spectrum.

We further produced an average asteroid spectrum for each collisional family, as well as for each mineralogical group of OC (H, L and LL). In the case of families, the spectra of individual family members were first reddened to match the spectral slope of the largest member. Next, the average spectrum was calculated at each wavelength, excluding data points diverging from the median value at the >2.5-σ level, and smoothed using a Savitzky-Golay filter. The spectral affinity between each family and each OC category was then measured by interpolating the average asteroid and meteorite spectra over a common wavelength grid with regular 0.005 µm step, and calculating the overall reduced $\chi^2$ statistics of the fit between the asteroid spectrum ($R_{ast}$) and OC spectra ($R_{met}$):

$$\chi^2 = 1/N \ \Sigma_i \ (R_{ast\_i} - R_{met\_i})^2 / \ \sigma_i^2,$$

where $N$ is the number of data points in the spectrum and $\sigma_i$ the standard deviation of the family members at wavelength $i$. The spectral contrast of the meteorite spectra were allowed to vary in order to match the band depth of the asteroid family spectrum. If three or fewer family members were observed (this is the case for the Juno and Maria families), then we assigned $\sigma_i$ = 0.05 at each wavelength. The spectral fits between the families and OCs are shown in SI Fig. 11.

Appendix C: Collisional evolution

Collisional evolution of the main belt and of the Massalia family was simulated with a statistical approach. Both populations were characterized by their SFDs, either cumulative $N(>D)$ or differential $dN(D)$. For each pair of SFDs $(i, j)$, and each pair of bins $(k, l)$ in each SFD, the numbers of collisions were estimated as

$$n_{ijkl} = p_{ij}\, f_g\, (D_{ik} + D_{jl})^2\, dN(D_{ik})\, dN(D_{jl})\, \Delta t,$$

where $p_{ij}$ is the collisional probability, $f_g$ the gravitational focussing factor, $D$ the diameters, $dN$ the discretized differential distributions, and $\Delta t$ the time step. We used the Monte-Carlo code 'Boulder' developed by Morbidelli et al. (2009) and subsequently substantially modified. A detailed description of our modifications is included in Brož et al. (2023). Here, we summarize the most important simulation parameters. We assumed the scaling law in the form of

$$Q^*(D) = 9.0 \times 10^7\ \text{erg g}^{-1}\, (D/2\ \text{cm})^{-0.53} + 0.5\ \text{erg cm}^{-3}\, \rho\, (D/2\ \text{cm})^{1.36},$$

where $Q^*$ is the specific energy necessary to disperse half of the target with the diameter $D$ and the density $\rho$. This was calibrated by the observed SFD of the main belt.

The decay time scale is also size-dependent, because it depends on the Yarkovsky effect, conductivity $K(D)$, spin rate $\omega(D)$, collisional reorientations, the YORP effect, the Poynting-Robertson effect, or determined by close encounters with planets (for NEOs). This was calibrated by the observed SFD of the NEOs.

The collisional probabilities and speeds for collisions between main belt-main belt, Massalia-main belt, and Massalia-Massalia were set to different values: $p_{ij}$ = 2.86, 4.26, and 29.00 × 10$^{-18}$ km$^{-2}$ y$^{-1}$, $v_{ij}$ = 5.77, 5.04, and 4.23 km/s, respectively. The latter determines the specific energy $Q = 1/2\, mv^2/M_{tot}$, consequently the ratio $Q/Q^*$ and the outcome — the largest remnant, the largest fragment, and the slope of all the fragments.

The observed SFD of the Massalia family has a cumulative slope with a power-law index of −5.7 between 5 and 2.5 km, and −2.8 between 1.2 and 0.8 km; a substantial bias due to observational incompleteness is present below approximately 0.5 km. In most families, a break at approximately 2.5 km is directly related to their collisional evolution (Brož et al. 2023), but such break is not seen in Massalia, which indicates a very young family age. For old families (e.g., Vesta, Flora), the SFD 'adapts' to the shallow SFD of the main belt, hence the break is shifted to subsequently larger and larger sizes.

The initial SFD of Massalia was set up as steep, with slope −5.7 down to 0.5 km size, but not more in order not to 'overshoot' the whole main belt. In the course of collisional evolution, the slope −5.7 naturally decreases to −2.8 over (450 ± 50) My. This is fully compatible with the timing of the L-chondrite event, but there are several problems: (i) the SFD is 'undershot' around 2.5 km; (ii) the current number of L chondrites is underestimated; (iii) the corresponding 1.4-deg dust band is not explained (see SI Fig. 7). In other words, it is impossible to keep an SFD steep for hundreds of My.

A number of different initial conditions were tested. For example, using an even steeper slope −6.5 would solve the problem (i), but not (ii) and (iii). The best-fit age of (800 ± 100) My is also too old to explain the L chondrites shock ages (Swindle et al. 2014). Using two slopes −5.7 and −2.8, close to the observed values, could result in a very young collisional age, perhaps comparable to previous estimates from the dynamics (Vokrouhlický et al. 2006), but this is in contradiction with the shock ages. An extreme SFD with slopes −5.7 and −3.0, which 'overshoots' the main belt, seems to be also compatible with observations, because the required parent body size ($D \sim 160$ km) is not unreasonable and because the collisional cascade immediately (in few My) decreases the respective population. In other words, the main belt is close to its own destruction by colliding with itself.

The only systematic solution to all three problems (i, ii and iii) seems to be that a second collisional event occurred on (20) Massalia. In order to build-up the whole SFD down to 100 µm, a sufficient ejection of material, with a volume-equivalent diameter of the order of 20 km, is needed. The only available body in the Massalia family is (20) Massalia itself, with $D = 131$ km; the second largest body measuring 5 km only. No sub-family is observed, because the ejection speed must be of the order of the escape speed in both cases. In other words, the fragment populations from the two collisions end up on similar orbital distributions, similarly to the case of Vesta. The resulting SFD is then a sum of two SFDs. The tail can be kept steep for up to 40 My (SI Fig. 7), in agreement with the distribution of cosmic-ray exposure ages of L chondrites (Eugster et al. 2006).

The first Massalia event is contemporary to a peak of cratering on Earth, composed of 20 craters dated 440 to 480 My, i.e., middle Ordovician (Schmieder & Kring 2020). Crater diameters are approximately 1 to 36 km; the largest being East Clearwater Lake, Canada. Assuming the projectile-to-crater ratio 1:20, this corresponds to projectile diameters 0.05 to 1.8 km. This is comparable with the Massalia family SFD, because it contains (contained) a number of bodies in this size range. However, the total number of impacts was probably too high and the probability of colliding with the Earth too low (0.3 %; Nesvorný et al. 2023) to be explained exclusively by the Massalia family.

Regarding the fossil L chondrites (Schmitz et al. 1997), the short duration of the event was previously attributed to the 5:2 resonance with Jupiter, which provides a fast-track route to the NEO space (Nesvorný et al. 2009). Here we propose instead that the duration is limited by the collisional cascade. If the first Massalia family 'overshot' the main belt by a factor of more than 10 – as required by the abundance of fossil L chondrites (e.g., Heck et al. 2017) – their collisional evolution inevitably leads to an immediate (few-My long) decrease of the population.

Regarding the excess extraterrestrial $^3$He found in sediments (Farley et al. 1998), its duration lasted from 34.4 to 36.2 My. Moreover, the magnitude of the excess is three times the background level. This is in agreement with our assumption that even the second Massalia family 'overshot' the whole main belt at micrometer sizes.

Appendix D: Orbital evolution

Orbital evolution of asteroids and meteoroids originating from the Massalia family was studied using N-body methods. We used the symplectic integrator RMVS3 from the SWIFT package (Levison and Duncan 1994), which was modified to include a computation of mean elements (Quinn et al. 1991), of proper elements (Šidlichovský and Nesvorný 1996), the Yarkovsky effect (Vokrouhlický 1998, Vokrouhlický and Farinella 1999), the YORP effect (Čapek and Vokrouhlický 2004), collisional reorientations (Farinella et al. 1998), and strength-dependent spin barrier (Holsapple 2007). A detailed description of our modifications was previously presented in Brož et al. (2011). Every simulation included 8 planets, the massive asteroids Ceres and Vesta, and approximately $10^3$ test particles. The time step was 9.13125 days, the output of osculating elements was performed every 10 ky, mean elements 3 ky, and proper elements 1 My.

For meter-sized bodies, the initial orbits corresponded to the observed family members, as identified by the hierarchical clustering (Brož et al. 2023). Their thermal parameters were set up as follows: a bulk density of 2500 kg m$^{-3}$, a thermal conductivity of 1 W m$^{-1}$ K$^{-1}$, a thermal capacity of 680 J kg$^{-1}$ K$^{-1}$, a Bond albedo of 0.1, an emissivity of 0.9, and a YORP efficiency of 0.33. The Yarkovsky drift rates were positive or negative, depending on spin axis orientations, |da/dt| < 0.15 au My$^{-1}$. We assumed the reorientation time scale in the form of

$\tau_{\rm reo}$ = 0.0845 My $(\omega/3.49 \times 10^{-4}$ rad s$^{-1})^{5/6}$ $(D/2$ m$)^{4/3}$.

The evolution of meteoroids is stochastic due to frequent reorientations. The chosen time span of 50 My is sufficient for the bodies to evolve across the inner belt and to enter the NEO region. Their lifetime depends on the pathway: the 3:1 mean-motion resonance with Jupiter at 2.5 au, the ν$_6$ secular resonance at approximately $i$ ~ 18 deg [(a/au − 2.1)/(2.5 − 2.1)]$^{-0.41}$, the 1:2 resonance with Mars at 2.4184 au, or diffusion through Mars-crossing region. We measured the decay time scale, $\tau_{\rm mb}$ = 139 My, in the main belt and the lifetime, $\tau_{\rm neo}$ = 3.83 My, in the NEO region. Assuming a steady state, the relation between the populations is determined by:

$N_{\rm neo}$(>1 m) = $N_{\rm mb}$(>1 m) $\tau_{\rm neo}/\tau_{\rm mb}$.

We also determined the inclination distribution in the NEO region (SI Fig. 9). The original inclination of the source family is partially preserved. In particular, it is difficult to scatter bodies by close encounters with planets from high to low inclinations. For the Massalia family (with ~1.4 deg), NEO orbits are scattered mostly to <13 deg. This is in agreement not only with the observed orbital distribution of L-chondrite-like NEOs (Fig. 3) but also with the pre-atmospheric orbits of L chondrites (SI Fig. 8).

SI References

[54] Rayner, J. T., Toomey, D. W., Onaka, P. M., Denault, A. J., Stahlberger, W. E., Vacca, W. D., Cushing, M. C., Wang, S., "SpeX: A Medium-Resolution 0.8-5.5 Micron Spectrograph and Imager for the NASA Infrared Telescope Facility", *Publications of the Astronomical Society of the Pacific*, vol. 115, no. 805, pp. 362–382, 2003. doi:10.1086/367745.

[55] Rivkin, A. S., Binzel, R. P., and Bus, S. J., "Constraining near-Earth object albedos using nearl-infrared spectroscopy", *Icarus*, vol. 175, no. 1, pp. 175–180, 2005. doi:10.1016/j.icarus.2004.11.005.

[56] Bus, S. J. and Binzel, R. P., "Phase II of the Small Main-Belt Asteroid Spectroscopic Survey. The Observations", *Icarus*, vol. 158, no. 1, pp. 106–145, 2002. doi:10.1006/icar.2002.6857.

[57] Burbine, T. H. and Binzel, R. P., "Small Main-Belt Asteroid Spectroscopic Survey in the Near-Infrared", *Icarus*, vol. 159, no. 2, pp. 468–499, 2002. doi:10.1006/icar.2002.6902.

[58] Britt, D. T., Pieters, C. M., "Black Ordinary Chondrites: an Analysis of Abundance and Fall Frequency", *Meteoritics*, vol. 26, no. 4, p. 279, 1991. doi:10.1111/j.1945-5100.1991.tb00727.x.

[59] Reddy, V., Sanchez, J. A., Bottke, W. F., Cloutis, E. A., Izawa, M. R. M., O'Brien, D. P., Mann, P., *et al.*, "Chelyabinsk meteorite explains unusual spectral properties of Baptistina Asteroid Family", *Icarus*, vol. 237, pp. 116–130, 2014. doi:10.1016/j.icarus.2014.04.027.

[60] Kohout, T., Gritsevich, M., Grokhovsky, V. I., Yakovlev, G. A., Haloda, J., Halodová, P., Michallik, R. M., *et al.*, "Mineralogy, reflectance spectra, and physical properties of the Chelyabinsk LL5 chondrite - Insight into shock-induced changes in asteroid regoliths", *Icarus*, vol. 228, pp. 78–85, 2014. doi:10.1016/j.icarus.2013.09.027.

[61] Kohout, T., Petrova, E. V., Yakovlev, G. A., Grokhovsky, V. I., Penttilä, A., Maturilli, A., Moreau, J.-G., *et al.*, "Experimental constraints on the ordinary chondrite shock darkening caused by asteroid collisions", *Astronomy and Astrophysics*, vol. 639, id. A146, 13 pp. 2020. doi:10.1051/0004-6361/202037593.

[62] DeMeo, F. E., Burt, B. J., Marsset, M., Polishook, D., Burbine, T. H., Carry, B., Binzel, R. P., *et al.*, "Connecting asteroids and meteorites with visible and near-infrared spectroscopy", *Icarus*, vol. 380, id. 114971, 2022. doi:10.1016/j.icarus.2022.114971.

[63] Cloutis, E. A., Gaffey, M. J., Jackowski, T. L., and Reed, K. L., "Calibrations of phase abundance, composition, and particle size distribution for olivine-orthopyroxene mixtures from reflectance spectra", *Journal of Geophysical Research*, vol. 91, pp. 11641–11653, 1986. doi:10.1029/JB091iB11p11641.

[64] Dunn, T. L., McCoy, T. J., Sunshine, J. M., and McSween, H. Y., "A coordinated spectral, mineralogical, and compositional study of ordinary chondrites", *Icarus*, vol. 208, no. 2, pp. 789–797, 2010. doi:10.1016/j.icarus.2010.02.016.

[65] Morbidelli, A., Bottke, W. F., Nesvorný, D., and Levison, H. F., "Asteroids were born big", *Icarus*, vol. 204, no. 2, pp. 558–573, 2009. doi:10.1016/j.icarus.2009.07.011.

[66] Vokrouhlický D., Brož, M., Bottke, W. F., Nesvorný, D., and Morbidelli, A., "Yarkovsky/YORP chronology of asteroid families", *Icarus*, vol. 182, no. 1, pp. 118–142, 2006. doi:10.1016/j.icarus.2005.12.010.

[67] Eugster, O., Herzog, G. F., Marti, K., and Caffee, M. W., "Irradiation Records, Cosmic-Ray Exposure Ages, and Transfer Times of Meteorites", *in* Meteorites and the Early Solar System II, D. S. Lauretta and H. Y. McSween Jr. (eds.), University of Arizona Press, Tucson, pp. 829–851, 2006.

[68] Nesvorný, D., Vokrouhlický, D., Morbidelli, A., and Bottke, W. F., "Asteroidal source of L chondrite meteorites", *Icarus*, vol. 200, no. 2, pp. 698–701, 2009. doi:10.1016/j.icarus.2008.12.016.

[69] Nesvorný, D. *et al.*, "NEOMOD: A New Orbital Distribution Model for Near Earth Objects", submitted, 2023.

[70] Farley K. A., Montanari, A., Shoemaker, E. M., and Shoemaker, C. S., "Geochemical Evidence for a Comet Shower in the Late Eocene", *Science*, vol. 280, no. 5367, p. 1250, 1998. doi:10.1126/science.280.5367.1250.

[71] Novaković, B., and Radović, V., "Asteroid Families Portal", EPSC-DPS Joint Meeting 2019, 2019. http://asteroids.matf.bg.ac.rs/fam/

[72] Levison, H. F., and Duncan, M. J., "The Long-Term Dynamical Behavior of Short-Period Comets", *Icarus*, vol. 108, no. 1, pp. 18–36, 1994. doi:10.1006/icar.1994.1039.

[73] Quinn, T. R., Tremaine, S., and Duncan, M., "A Three Million Year Integration of the Earth's Orbit", *Astronomical Journal*, vol. 101, p. 2287, 1991. doi:10.1086/115850.

[74] Šidlichovský, M. and Nesvorný, D. "Frequency modified Fourier transform and its applications to asteroids", *Celestial Mechanics and Dynamical Astronomy*, vol. 65, no. 1–2, pp. 137–148, 1996. doi:10.1007/BF00048443.

[75] Vokrouhlický, D., "Diurnal Yarkovsky effect as a source of mobility of meter-sized asteroidal fragments. I. Linear theory", *Astronomy and Astrophysics*, vol. 335, pp. 1093–1100, 1998.

[76] Vokrouhlický, D., and Farinella, P., "The Yarkovsky Seasonal Effect on Asteroidal Fragments: A Nonlinearized Theory for Spherical Bodies", *Astronomical Journal*, vol. 118, no. 6, pp. 3049–3060, 1999. doi:10.1086/301138.

[77] Čapek, D., and Vokrouhlický, D., "The YORP effect with finite thermal conductivity", *Icarus*, vol. 172, no. 2, pp. 526–536, 2004. doi:10.1016/j.icarus.2004.07.003.

[78] Farinella, P., Vokrouhlický, D., and Hartmann, W. K., "Meteorite Delivery via Yarkovsky Orbital Drift", *Icarus*, vol. 132, no. 2, pp. 378–387, 1998. doi:10.1006/icar.1997.5872.

[79] Holsapple, K. A., "Spin limits of Solar System bodies: From the small fast-rotators to 2003 EL61", *Icarus*, vol. 187, no. 2, pp. 500–509, 2007. doi:10.1016/j.icarus.2006.08.012.

[80] Brož, M., Vokrouhlický, D., Morbidelli, A., Nesvorný, D., and Bottke, W. F., "Did the Hilda collisional family form during the late heavy bombardment?", *Monthly Notices of the Royal Astronomical Society*, vol. 414, no. 3, pp. 2716–2727. 2011. doi:10.1111/j.1365-2966.2011.18587.x.